\documentclass[12pt,nofootinbib,tightenlines,preprintnumbers,superscriptaddress]{revtex4}
\usepackage[T1]{fontenc}
\usepackage[utf8]{inputenc}
\usepackage{lmodern}
\usepackage[english]{babel}
\usepackage{graphicx}
\usepackage{amsmath,mathtools,amssymb}
\usepackage{hyperref}
\usepackage{tabularx}
\usepackage{hhline}
\usepackage[dvipsnames]{xcolor}
\usepackage{tikz}
\usetikzlibrary{decorations.pathreplacing,decorations.pathmorphing}
\usepackage{tikz-feynman}

\renewcommand{\vec}[1]{\boldsymbol{#1}}
\newcommand{\be}{\begin{equation}}
\newcommand{\ee}{\end{equation}}
\newcommand{\ba}{\begin{eqnarray}}
\newcommand{\ea}{\end{eqnarray}}
\newcommand{\la}{\label}

\newcommand{\<}{\langle}
\renewcommand{\>}{\rangle}

\newcommand{\txts}{\textstyle}

\newcommand{\dint}{\mathrm d}
\usepackage{physics}

\begin{document}

\title{Deep inelastic scattering on the quark-gluon plasma}

\preprint{MITP/20-075, CERN-TH-2020-206}
      
\author{Marco C{\`e}} 
\affiliation{Theoretical Physics Department, 
CERN, 
CH-1211 Geneva 23, Switzerland}

\author{Tim Harris} 
\affiliation{Dipartimento di Fisica, 
Universit{\`a} di Milano-Bicocca, and INFN, sezione di Milano-Bicocca, 
Piazza della Scienza 3, I-20126 Milano, Italy}

\author{Harvey B.\ Meyer} 
\affiliation{Helmholtz~Institut~Mainz,
Johannes Gutenberg-Universit\"at Mainz,
D-55099 Mainz, Germany}
\affiliation{PRISMA$^+$ Cluster of Excellence \& Institut f\"ur Kernphysik,
Johannes Gutenberg-Universit\"at Mainz,
D-55099 Mainz, Germany}

\author{Arianna Toniato} 
\affiliation{PRISMA$^+$ Cluster of Excellence \& Institut f\"ur Kernphysik,
Johannes Gutenberg-Universit\"at Mainz,
D-55099 Mainz, Germany}

\begin{abstract}
We provide an interpretation of the structure functions of a thermal medium
such as the quark-gluon plasma in terms of the scattering of an incoming electron 
on the medium via the exchange of a spacelike photon. 
We then focus on the deep-inelastic scattering (DIS) regime, and formulate the 
corresponding moment sum rules obeyed by the structure functions. Accordingly, these moments
are given by the thermal expectation value of twist-two operators, which is 
computable from first principles in lattice QCD for the first few moments.
We also show how lattice QCD calculations can be used to probe how large the photon virtuality
needs to be in order for the Bjorken scaling of structure functions to set in.
Finally, we provide the parton-model interpretation of the structure
functions in the Bjorken limit and test its consistency.  As in DIS on the proton, the kinematic 
variable $x$ is proportional to the longitudinal momentum carried by the
partons, however $x$ ranges from zero to infinity. Choosing the parton momentum parametrization to be  $ x T u$
where $u$ is the fluid four-velocity and $T$ its temperature in the rest frame, 
the parton distribution function for a plasma of non-interacting quarks is proportional to $ x \log(1+e^{-x/2}) $.
\end{abstract}

\maketitle

\section{Introduction}

Strong-interaction matter at high temperature is investigated
extensively in high-energy heavy-ion
collisions~\cite{Braun-Munzinger:2015hba,Busza:2018rrf}.  The
time-evolution of the matter produced in these collisions has been
described successfully using relativistic
hydrodynamics~\cite{Ollitrault:2007du,Teaney:2009qa}. Although the
reaction occurs extremely rapidly, the system appears to reach an
approximate local thermal equilibrium.
The typical temperatures achieved are around 300\,MeV. 
The value of the shear viscosity to entropy density ratio (around
0.20~\cite{Everett:2020yty}) extracted from comparing observed soft
particle spectra with the outcome of the hydrodynamics calculations is
the lowest of all known forms of matter, indicating that the produced
matter behaves like an excellent fluid.

The high-temperature phase of QCD is known as the quark-gluon plasma
(QGP).  At extremely high temperatures, due to asymptotic
freedom, one indeed expects the thermodynamic and transport properties of the
medium to be dominated by quasiparticles carrying the quantum numbers
of either quarks or gluons. Whether such a simple picture applies even
approximately at the temperatures reached in heavy-ion collisions is
questionable in view of the phenomenological findings described
above. If one then accepts the picture of a strongly correlated fluid
for the QGP at $T=300\,$MeV, a new question arises. Probed on short
enough distance and time scales, surely this fluid must exhibit quark
and gluon scatterers; this property is guaranteed by asymptotic
freedom and is analogous to the statement that quarks and gluons can
be `seen' inside the proton when probed in deep inelastic
scattering. How then do the fluid properties of the QGP emerge from
these effective degrees of freedom seen on short time scales?  This
question provides part of the motivation behind the present work.  It
has recently been addressed \cite{DEramo:2018eoy} by invoking the
large-angle scattering of jets or parts thereof on the QGP in a
heavy-ion collision. Here we invoke the idealised situation of a light
lepton scattering on the QGP via the exchange of a spacelike photon,
and ask the question in reverse: starting from the fluid in thermal equilibrium,
at what resolution scale do the quarks become `visible'?
Although it appears unlikely that such scattering could be identified
in a heavy-ion collision, the interaction of a lepton with the QGP is
conceptually simpler, and it turns out that it can be probed in
lattice QCD via fixed-virtuality dispersion relations.  At large
photon virtuality $Q^2$, the cross-section per unit volume can be understood
in the framework of deep inelastic scattering (DIS). The structure
functions of the medium then depend only logarithmically on $Q^2$, and
a parton distribution function of the QGP can be defined.

There is a further motivation for studying the structure functions of
the QGP.  The latter are connected via a Kubo-Martin-Schwinger
relation to the spectral functions through which the thermal
production rate of photons and dileptons are usually expressed. These rates
are of central importance for understanding the corresponding spectra
measured in heavy-ion
collisions~\cite{Braun-Munzinger:2015hba,David:2019wpt}.  As far as
the photon emissivity is concerned, it has recently been shown that it
can be probed in lattice QCD at fixed, vanishing virtuality
$Q^2=0$~\cite{Meyer:2018xpt}.  Probing the dilepton rate, however, is
only possible via a dispersion relation at fixed spatial momentum
$\vec q$ of the lepton pair; see for instance
references~\cite{Ding:2010ga,Brandt:2012jc,Amato:2013naa,Aarts:2014nba,Brandt:2015aqk,Ding:2016hua}.
In these calculations, the
Euclidean correlator probes both the spacelike regime of the spectral
function and the timelike regime, since it is represented as an
integral over all positive real frequencies $q^0$.  For that reason,
it turns out that understanding the behaviour of the spectral
functions in the DIS regime can be helpful in probing the low-mass
dilepton production rate with a large spatial momentum relative to the
plasma in lattice QCD. We will return to this point at the end of
section \ref{sec:lepton}.

We note that certain aspects of DIS on non-Abelian plasmas have been
addressed previously in the references \cite{Hatta:2007cs,Iancu:2009zz}.
In the first one, DIS on an ${\cal N}=4$ super Yang-Mills plasma
is investigated using the AdS/CFT correspondence,
and the possibility of interpreting the corresponding structure functions
in terms of partons is discussed in detail. The impact-parameter representation
plays an important role in the analysis.
The second reference proposes to evaluate the thermal expectation value of a
particular twist-two operator in QCD, and more concretely in quenched QCD,
in order to judge whether its size is more typical of a weak-coupling
or a strong-coupling `scenario'. Its relation to the medium's structure functions
via a moment sum rule is used to infer a partonic interpretation of these two
scenarios.
With respect to these references, our analysis differs by its emphasis on full QCD
and by technical aspects; in particular the range of the
Bjorken-$x$ variable is different, the fixed-virtuality dispersion
relations are new and the connection to the spectral functions
discussed in the context of photon and dilepton production plays a
major role. This paper also has common aspects with the work
of Ref.~\cite{Chambers:2017dov}, which studies the nucleon structure
functions on the lattice via fixed-virtuality dispersion relations.
A difference with the nucleon structure functions and, more generally,
with zero-temperature structure functions, is that in the thermal case
one cannot simply take derivatives of Euclidean correlators with respect to the variable $q^0$
without confronting a numerically ill-posed analytic continuation problem.
This difference stems from the fact that the thermal correlator
can be computed only at the discrete Matsubara frequencies,
$q^0=i 2\pi Tn$, $n\in\mathbb{Z}$.

The rest of this paper is structured as follows.
Section~\ref{sec:lepton} presents the derivation of the cross-section
per unit volume for a lepton scattering on a thermal medium in terms
of its structure functions, and provides their connection to the
spectral functions. The Bjorken limit corresponding to deep-inelastic
scattering is formulated and the structure functions for a plasma of
non-interacting quarks are given in that limit.  In
section~\ref{sec:sr}, the moment sum rules are presented for the
medium structure functions, allowing the first few moments to be
computed non-perturbatively in lattice QCD.  Particular attention is
devoted to the $n=2$ sum rule, corresponding to the momentum sum rule
in the parton model.  Opportunities for lattice QCD studies of the
medium structure functions for spacelike momenta are presented in
section \ref{sec:fixedQ2}. Finally, the parton-model interpretation of
the structure functions in the DIS limit is discussed in
section~\ref{sec:parton}.  We summarize and conclude in
section~\ref{sec:concl}.  A number of technical aspects are collected
in appendix, in particular the derivation of the moment sum rules and
the explicit verification of the first two sum rules for the plasma of
non-interacting quarks.

\section{Lepton scattering on a thermal medium and structure functions of the latter\la{sec:lepton}}

In this section, we recall some of the basic aspects of the inelastic
scattering of a lepton on hadronic matter. Our notation is largely
taken from the lecture notes~\cite{Manohar:1992tz}.  Let $k$ be the
initial momentum of the lepton, and $k'$ its final momentum.  We also
use $E$ and $E'$ respectively to denote the zeroth components of these
vectors. The momentum transferred to the medium is thus $q=k-k'$, and
we use $Q^2 = -q^2 \geq 0 $ to denote the spacelike virtuality of the
exchanged photon. The lepton mass is neglected throughout. 

\begin{figure}[t]
  \begin{tikzpicture}[baseline=0]
    \pgfmathsetmacro{\ix}{-3};
    \pgfmathsetmacro{\iy}{-1};
    \pgfmathsetmacro{\l}{0.4};
    \pgfmathsetmacro{\dx}{0.1};
    \pgfmathsetmacro{\dy}{0.1};
    \begin{scope}[xshift=-22.5mm,yshift=-25mm,xslant=-1,yscale=0.25]
      \coordinate (A) at (0,0);
      \coordinate (B) at (6,0);
      \coordinate (C) at (6,6);
      \coordinate (D) at (0,6);
      \draw (A) rectangle (C) node[midway] {QGP};
    \end{scope}
    \draw (A) -- ++(0,1.7); \draw[dotted] (A) ++(0,1.7) -- ++(0,2.5);
    \draw (B) -- ++(0,4.5);
    \draw (C) -- ++(0,0.2); \draw[dotted] (C) ++(0,0.2) -- ++(0,3.5);
    \draw (D) -- ++(0,3.5);
    \begin{feynman}
      \vertex (i1) at (-3.0, 2.8) {$e^-$};
      \vertex (o1) at (+3.0, 1.8) {$e^-$};
      \vertex (v1) at ( 0.5, 1.8);
      \vertex (v2) at ( 0.0, 0.0);
      \diagram* {
        (i1) -- [fermion,momentum=$k$] (v1) -- [fermion,momentum=$k'$] (o1),
        (v1) -- [boson, edge label'=$\gamma$,momentum=$q$] (v2),
      };
    \end{feynman}
    \draw[-Stealth] (1.2,-2.2) -- (2.2,-2.2) node[midway,above] {$u$};
    \draw[fill=white] (\ix-\l/2,-\l/2) rectangle ++(\l,\l) node[midway] {\footnotesize $\phantom{\Omega}$} -- ++(-\dx,\dy) -- ++(-\l,0) -- ++(\dx,-\dy) ++(0,0) -- ++(-\dx,\dy) -- ++(0,-\l) -- ++(\dx,-\dy);
    \draw[very thin,rotate around={ 2:(\ix/2,0)}] (0.85*\ix,-0.4) -- (0.15*\ix,-0.4);
    \draw[very thin,rotate around={ 3:(\ix/2,0)}] (0.85*\ix,-0.2) -- (0.15*\ix,-0.2);
    \draw[very thin,rotate around={-1:(\ix/2,0)}] (0.85*\ix, 0)   -- (0.15*\ix, 0);
    \draw[very thin,rotate around={ 1:(\ix/2,0)}] (0.85*\ix,+0.2) -- (0.15*\ix,+0.2);
    \draw[very thin,rotate around={ 0:(\ix/2,0)}] (0.85*\ix,+0.4) -- (0.15*\ix,+0.4);
    \draw[fill=white] (-\l/2,-\l/2) rectangle ++(\l,\l) node[midway] {\footnotesize $\phantom{\Omega}$} -- ++(-\dx,\dy) -- ++(-\l,0) -- ++(\dx,-\dy) ++(0,0) -- ++(-\dx,\dy) -- ++(0,-\l) -- ++(\dx,-\dy);
    \draw[very thin,rotate around={ 2:(-\ix/2,0)}] (-0.15*\ix,-0.4) -- (-0.85*\ix,-0.4);
    \draw[very thin,rotate around={ 3:(-\ix/2,0)}] (-0.15*\ix,-0.2) -- (-0.85*\ix,-0.2);
    \draw[very thin,rotate around={-1:(-\ix/2,0)}] (-0.15*\ix, 0)   -- (-0.85*\ix,-0.9);
    \draw[very thin,rotate around={ 1:(-\ix/2,0)}] (-0.15*\ix,+0.2) -- (-0.85*\ix,+0.2);
    \draw[very thin,rotate around={ 0:(-\ix/2,0)}] (-0.15*\ix,+0.4) -- (-0.85*\ix,+0.4);
    \draw[decorate,decoration=brace] (-0.9*\ix,0.5) -- (-0.9*\ix,-1.1) node[midway,right] {$X$};
  \end{tikzpicture}\qquad\qquad
  \begin{minipage}{0.4\textwidth}
    \vspace{0.5cm}
  \begin{tikzpicture}[baseline=0]
    \begin{feynman}
      \vertex (i1) at (-3.0, 2.8) {$e^-$};
      \vertex (o1) at (+3.0, 1.8) {$e^-$};
      \vertex (v1) at ( 0.5, 1.8);
      \vertex (i2) at (-3.0, 0.0) {${\rm q}$};
      \vertex (o2) at (+3.0,-1.0) {${\rm q}$};
      \vertex (v2) at ( 0.0, 0.0);
      \diagram* {
        (i1) -- [fermion,momentum=$k$]  (v1) -- [fermion,momentum=$k'$]  (o1),
        (i2) -- [fermion,momentum'=$p_\xi$] (v2) -- [fermion,momentum'=$p_\xi'$] (o2),
        (v1) -- [boson, edge label'=$\gamma$,momentum=$q$] (v2),
      };
    \end{feynman}
  \end{tikzpicture}
  \end{minipage}
  \bigskip
  
  \caption{Left: Scattering of a lepton on quark-gluon plasma at global thermal equilibrium moving with a four-velocity $u$ in the one-photon exchange approximation.
    The picture shows the interaction occurring with a cubic fluid cell, producing an unobserved QCD final-state $X$.
Right: interpretation of the process in the deep-inelastic scattering limit as an elastic parton-lepton collision.
    \la{fig:kin}}
\end{figure}
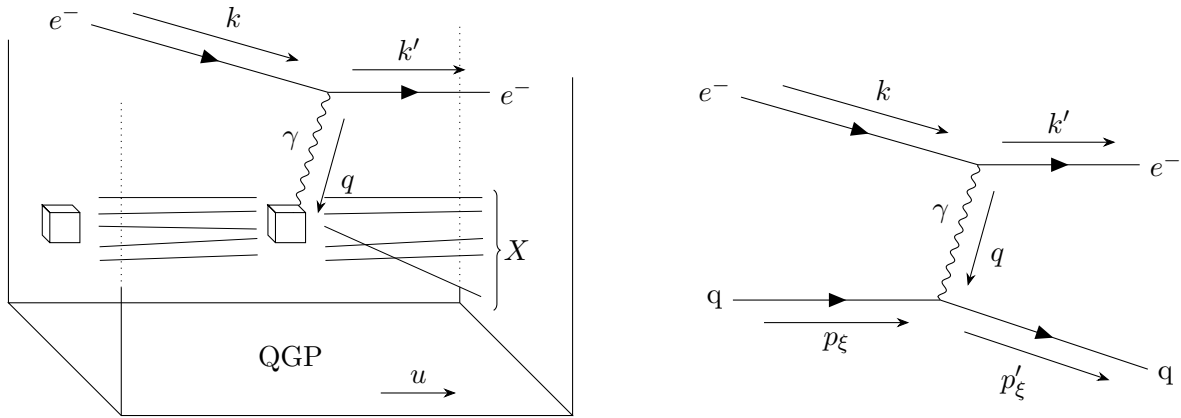

\subsection{Inelastic cross section and hadronic tensor}

First, recall the expression for the inclusive differential cross section for lepton-proton scattering,
\ba\la{eq:dsigma}
 d\sigma &=& \sum_X \int \frac{d^3k'}{(2\pi)^3 2E'}\; \frac{(2\pi)^4 \delta^4(k+p-k'-p_X)}{(2E)(2M)} \,\frac{e^4}{Q^4}\,
 \\ &&
\qquad \qquad \qquad \qquad \times  \<p,\lambda|j^\mu(0)|X\>\;\<X|j^\nu(0)|p,\lambda\>\;\ell_{\mu\nu}\;, 
\phantom{\frac{1}{1}}
\nonumber
\\ \ell_{\mu\nu} &=& \sum_{s_l'} \<k,s_l|j_{\ell\mu}(0)|k',s_l'\>\,\<k',s_l'|j_{\ell\nu}(0)|k,s_l\>\;,
\ea
where $M$ is the proton mass and the proton-state normalisation 
is $\<p'\lambda'|p\lambda\> = 2E_{\vec p} (2\pi)^3 \delta(\vec p-\vec p') \delta_{\lambda\lambda'}$. 
The leptonic tensor $\ell_{\mu\nu}$ is easily computed, and its expression can be found in~\cite{Manohar:1992tz}.
With
\be
j^\mu(x) = \sum_{f=u,d,s,\dots} {\cal Q}_f\; \bar\psi_f(x) \gamma^\mu \psi_f(x),
\ee
the electromagnetic current carried by the quarks\footnote{The physical values of the quark charges are
${\cal Q}_u = {\txts\frac{2}{3}},~ {\cal Q}_d={\cal Q}_s = -{\txts\frac{1}{3}},~\dots$}, one definition of the hadronic tensor is given by 
\ba
W_>^{\mu\nu}(p,q)_{\lambda'\lambda} &=& \frac{1}{4\pi}\int d^4x\; e^{iq\cdot x}\,\<p,\lambda'|j^\mu(x)\,j^\nu(0)|p,\lambda\>
\\ &=& \frac{1}{4\pi} \sum_X (2\pi)^4  \delta^4(q+p-p_X)\; \<p,\lambda'|j^\mu(0)|X\>\;\<X|j^\nu(0)|p,\lambda\>,
\nonumber
\ea
which allows one to substitute the sum over final states $X$ by $W_>^{\mu\nu}(p,q)$. In fact, 
the hadronic tensor $W^{\mu\nu}(p,q)$ is usually defined by replacing the product of currents by their commutator.
The tensors $W_>^{\mu\nu}$ and $W^{\mu\nu}$ are equal for the kinematics accessible in $ep$ scattering.
One then obtains for the differential cross section
\be
\frac{d^2\sigma}{dE'd\Omega} = \frac{e^4}{8\pi^2Q^4} \left(\frac{E'}{ME}\right) \ell_{\mu\nu} \;\frac{1}{2}\sum_{\lambda} W^{\mu\nu}(p,q)_{\lambda\lambda}.
\ee

The derivation of the cross-section for a lepton scattering on thermal hadronic matter is very similar to the case of scattering on a proton.
See the left panel of figure \ref{fig:kin} illustrating the process.
If $L^3$ is the volume of the thermal system in its rest frame, the differential cross section reads
\be
\frac{d^2\sigma}{dE'd\Omega} = \frac{e^4\,L^3}{8\pi^2Q^4} \left(\frac{E'}{E}\right) \ell_{\mu\nu}W_>^{\mu\nu}(u,q),
\ee
where the hadronic tensor is given by
\be
W_>^{\mu\nu}(u,q) = \frac{1}{4\pi\,Z} \sum_n e^{-\beta E_n} \int d^4x\; e^{iq\cdot x}\,\<n|j^\mu(x)\,j^\nu(0)|n\>,
\ee
 $Z= \sum_n e^{-\beta E_n}$ being the canonical partition function.
While the temperature dependence of $W_>^{\mu\nu}$ is not indicated explicitly,
the argument $u^\mu$ represents the four-velocity of the fluid. The state $|n\>$ is unit-normalized.
In inelastic scattering on the proton, $q^0=E-E'\geq0$, since the energy of the final hadronic state can only be larger than the proton mass;
in scattering on the plasma, it is possible for the electron to gain energy ($q^0<0$), although this is suppressed as $e^{-\beta |q^0|}$.
Here, due to our normalisation of the $|n\>$ states, the $(2M)^{-1}$ factor appearing in Eq.\ (\ref{eq:dsigma}) should not be 
included; instead a factor $L^3$ should be included.
We recall that for a thermal state, the typical energy of a contributing state $|n\>$ is of order volume, while its spatial momentum is of order $L^{3/2}$
in the canonical ensemble (see e.g.\cite{Giusti:2010bb}). Therefore, working in the thermodynamic limit and in the rest frame of the fluid, 
a typical state $|n\>$ can be treated as if it were at rest.

Through the Kubo-Martin-Schwinger relation, $W_>^{\mu\nu}$ can be rewritten in terms of the spectral function $\rho^{\mu\nu}$, 
which is defined as the Fourier transform of the commutator of the two currents, analogously to $W^{\mu\nu}$ in the proton case
(see Eq.\ \ref{eq:rho_def} for the explicit definition).
Using Eq.\ (38) of Ref.~\cite{Meyer:2011gj}\footnote{Our present convention for the normalization of the spectral function is that no factor (of $1/(4\pi)$ or other)
appears in front of the Fourier transform of the current commutator, 
which makes the spectral function defined here a factor $(2\pi)$ larger than in the convention used in~\cite{Meyer:2011gj}.},
\be
W_>^{\mu\nu}(u,q) = \frac{1}{4\pi(1-e^{-\beta q^0})}\,\rho^{\mu\nu}(q^0,\vec q) \qquad \textrm{(fluid rest frame).}
\ee
It is worth pointing out here that in the DIS regime defined below, since $q^0$ is extremely large,  the structure function only differs from 
the spectral function by the purely conventional factor of $4\pi$.
One may parametrize the tensor $W_>^{\mu\nu}$ in a way similar to the proton case,
\be\la{eq:Wmunu_decomp}
W_>^{\mu\nu}(u,q) = F_1(u\cdot q,Q^2) \Big(-g^{\mu\nu} + \frac{q^\mu q^\nu}{q^2}\Big) + \frac{m_T}{u\cdot q} F_2(u\cdot q,Q^2)
\Big(u^\mu-(u\cdot q) \frac{q^\mu}{q^2}\Big) \Big(u^\nu-(u\cdot q) \frac{q^\nu}{q^2}\Big).
\ee
Here $m_T$ is some thermal energy scale; we will later choose to set $m_T=T$.
The contraction of $W_>^{\mu\nu}$ with the symmetric part of the leptonic tensor
\be
\ell^{\{\mu\nu\}} = 2(k^\mu k'{}^\nu + k^\nu k'{}^\mu - g^{\mu\nu} (k\cdot k'))
\ee
yields
\be
\ell_{\{\mu\nu\}} W_>^{\mu\nu}= \frac{4 m_T (u\cdot k)}{y} \,\Big(x y^2 F_1 + (1-y) F_2\Big) \, ,
\ee
where
\be\la{eq:x.and.y}
x = \frac{Q^2}{2 m_T (u\cdot q)}, \qquad \qquad y = \frac{u\cdot q}{u\cdot k} \leq 1 \, .
\ee
The last inequality comes from the requirement that the lepton cannot lose more than its initial energy $E$.
We note that the definition of $x$ is analogous to the definition of Bjorken-$x$ in DIS on the proton.
One qualitative difference between scattering on the proton and on the fluid is that in the latter case, the value $x=1$
does not have any special significance, while for the proton it corresponds to elastic scattering.
Thus the multiplicative normalization of $x$ has no particular meaning, which
is reflected in the freedom of choosing the precise value of $m_T$.
For convenience, we may introduce 
\be
\rho_{1,2}(q^0,Q^2) = 4\pi (1-e^{-\beta q^0}) F_{1,2}(q^0,Q^2),
\qquad \textrm{(fluid rest frame)}
\label{eq:rho12}
\ee
so that the cross-section may finally be written in terms of the spectral functions,
\be\la{eq:dsigdEpdOm}
\frac{d^2\sigma}{L^3 dE'd\Omega} =  \frac{2\alpha^2 \,E' m_T }{\pi Q^4 y(1-e^{-\beta q^0})} \,
\Big(xy^2 \rho_1(q^0,Q^2) + (1-y) \rho_2(q^0,Q^2)\Big),
\qquad \textrm{(fluid rest frame)}.
\ee
One easily verifies that $\rho_1/q^0$ and $\rho_2$ are even, non-negative functions
of $q^0$ at fixed $|\vec q|$~\cite{Meyer:2011gj}, and hence at fixed
$Q^2\geq 0$.  Therefore the cross-section (\ref{eq:dsigdEpdOm}) is
guaranteed to be positive for all values of $y\leq 1$.  We recall that
the DIS limit corresponds to $Q^2\to +\infty$ with the variable $x$
fixed.  In this limit, due to the asymptotic freedom property of QCD,
the structure functions $F_{1,2}$ and the corresponding spectral
functions $\rho_{1,2}$ are expected to depend mainly on $x$, and only
logarithmically on $Q^2$.  In the parton-model interpretation of the
process, the lepton scattering cross section arises from the
incoherent sum of lepton-parton scatterings; see the right panel of figure \ref{fig:kin}. Given that in QCD 
these partons are massless spin-1/2 particles, this picture implies
the Callan-Gross relation, $F_2 = 2xF_1 $, which receives corrections logarithmically suppressed in $Q^2$.

We set $m_T=T$ from here on until section (\ref{sec:parton}), in which
the role of this thermal energy scale is discussed within the parton model.

\subsection{Structure functions of the non-interacting plasma}

Working in the rest frame of the medium,
we note that spectral functions are usually calculated as a function of $q^0$ (also often notated $\omega$) and $\vec q$.
We remark that if one chooses $|\vec q|$ and $x$ as independent variables, the frequency variable takes the form
\be\la{eq:q0DIS}
 q^0  =  \sqrt{\vec q^2 + T^2 x^2} \;- Tx \stackrel{|\vec q|\to\infty}{\simeq}  |\vec q| - T x.
\ee
The DIS regime can be reached by sending $|\vec q|\to\infty$ at fixed $x$,  since $Q^2= 2T|\vec q|x +{\rm O}(T^2)$.

Thermal QGP spectral functions at leading order in perturbation theory are given for instance in~\cite{Laine:2013vma}.
Taking the leading-order expression of $\rho_1$, which is equal to $\rho^{11}$ in the fluid  rest frame for $q^\mu=(q^0,0,0,q^3)$,
one finds, for $N_c$ colors of massless quarks,
\be
\lim_{Q^2\to\infty} F_1(x, Q^2) = 
\frac{ (\sum_f \mathcal Q_f^2) N_cT^2}{4\pi^2} x \log(1+e^{-x/2}).
\label{eq:freeF1}
\ee
One also finds that $-\frac{1}{2}W_>^\mu{}_\mu  $ has the same limit. It follows that 
\be
\lim_{Q^2\to\infty} (F_2 - 2x F_1) = 0,
\ee
which is the Callan-Gross relation.
Thus the leading-order calculation of the spectral functions confirms that the $Q^2$ dependence disappears in the Bjorken limit,
and that the Callan-Gross relation is satisfied.

\begin{figure}
\includegraphics[width = .8\textwidth]{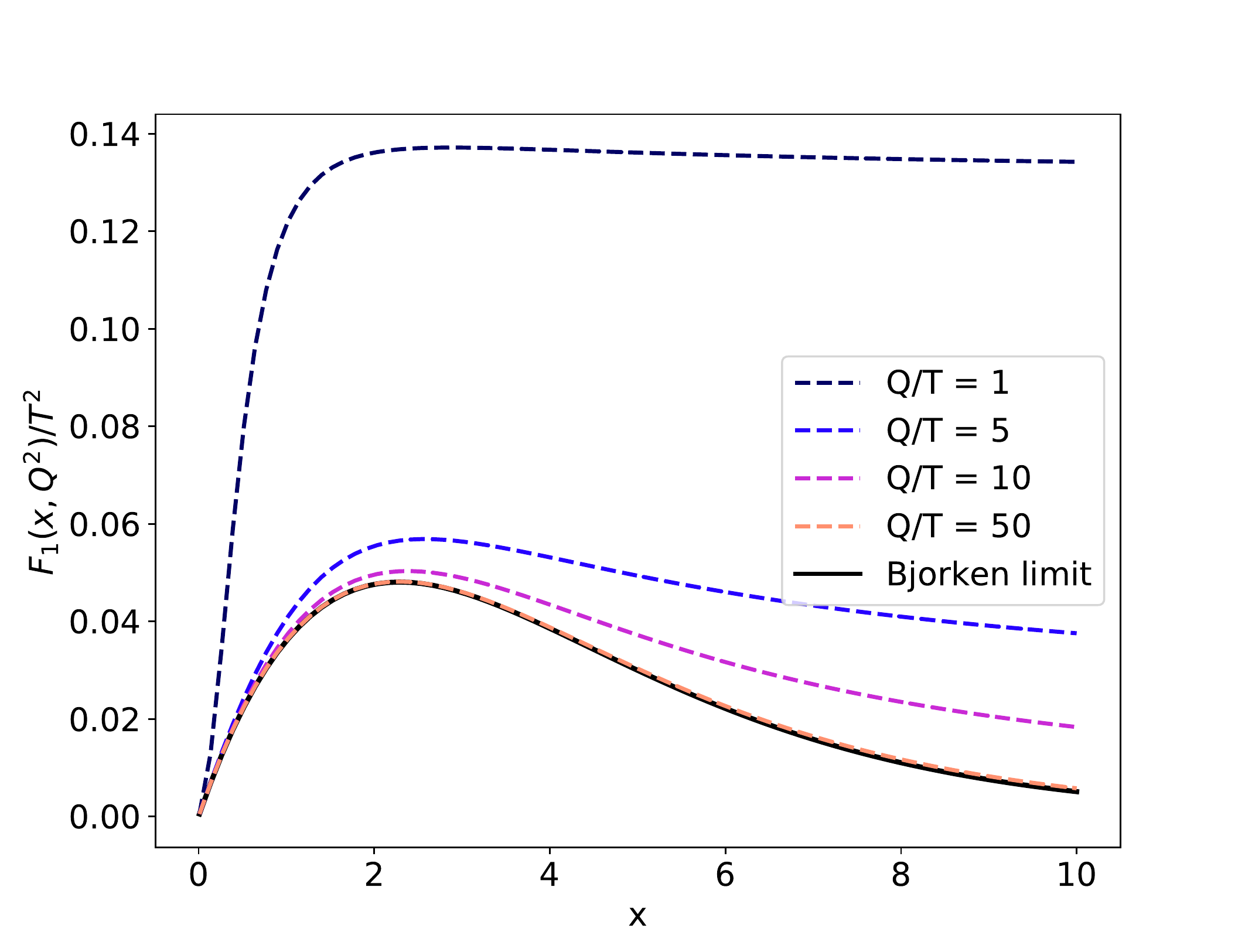}
\caption{The structure function $F_1(x,Q^2)$ of the free plasma for different values of $Q^2$, together with the Bjorken limit Eq.~(\ref{eq:freeF1}).
  Here we have set $N_c=3$ and $\sum_f {\cal Q}_f^2=1$.
}
\label{fig:Bjlim_free}
\end{figure}

The graph of expression (\ref{eq:freeF1}) is displayed in
Fig.~\ref{fig:Bjlim_free}. It is positive and normalizable as a
function of $x$.  The approach of $F_1$ to the expression in
Eq.\ (\ref{eq:freeF1}) is also illustrated the figure. Clearly, $F_1$
approaches its limit sooner at small $x$ than at large $x$. It is not before 
$Q/T=10$ that $F_1$ qualitatively acquires its asymptotic shape.

We now briefly return to a remark made in the introduction. As noted around Eq.\ (\ref{eq:q0DIS}),
the DIS regime can be reached by setting $q^0 = |\vec q|- xT$ for a fixed
value of the Bjorken variable $x$, and sending $|\vec q|\to\infty$.
Thus, in a dispersive representation at large and fixed $|\vec q|$, the Euclidean correlator
receives a contribution from the spectral function in the DIS regime.
More precisely, if $Q_0^2$ denotes the virtuality beyond which the DIS
regime applies for $x$ of order unity, the range of frequencies corresponding to the DIS regime is
\be
|\vec q | -\pi T \lesssim q^0 \leq |\vec q| - \frac{Q_0^2}{2|\vec q|}.
\ee
Therefore, understanding the behaviour of the spectral functions in
the DIS regime is helpful for probing the low-mass dilepton production
rate with a large spatial momentum relative to the plasma using lattice
QCD correlators in conjunction with a fixed-$|\vec q|$ dispersion relation.

\section{Sum rules\la{sec:sr}}

A standard tool to analyze the structure functions in the DIS regime is the operator-product expansion (OPE).
Here we review the OPE analysis to formulate the moment sum rules of the structure functions.
We set $N_c=3$ in this section, as appropriate for QCD.

The time-ordered operator-product of two electromagnetic current admits the asymptotic expansion
\begin{equation}
\begin{split}
t_{\mu\nu} (q) &  \equiv \: i \int \dint^4 x \: e^{i q \cdot x}\: {\rm T} \{ j_{\mu} (x) j_{\nu} (0) \} \\
& \overset{Q \to \infty}{\sim} \sum_{n=2,4,\dots} \sum_{f,j} c_{f;\mu \nu \mu_1 \dots \mu_n} (q) \: M_{fj} (Q, \tilde\mu) \:
\bigl[ O_{nj}^{\mu_1 \dots \mu_n} \bigr]_{\tilde\mu} \: ,
\end{split}
\label{eq:OPE}
\end{equation}
where $\tilde\mu$ is a reference energy scale at which the local operators on the right-hand side are renormalized.
The twist-two fermionic operator with flavor $f$ is 
\begin{equation}\la{eq:Onf_def}
O_{nf}^{\mu_1 \dots \mu_n} = \frac{1}{2} \biggl( \frac{i}{2} \biggr)^{n-1} 
\mathcal S \{ \bar\psi_f \gamma^{\mu_1} \overleftrightarrow{D}^{\mu_2} \dots \overleftrightarrow{D}^{\mu_{n}}\psi_f \}
\: ,
\end{equation}
and the gluonic one reads
\begin{equation}
O_{ng}^{\mu_1 \dots \mu_n} = - \frac{1}{2} \biggl( \frac{i}{2} \biggr)^{n-2} 
\mathcal S \{ F^{\mu_1 \alpha} \overleftrightarrow{D}^{\mu_2} \dots \overleftrightarrow{D}^{\mu_{n-1}} F^{\mu_n}_{\quad \alpha} \}
\: .
\end{equation}
Here $\overleftrightarrow{D} = \overrightarrow{D}-\overleftarrow{D}$ and ${\cal S}$ 
acts as $1/n!$ times the sum over permutations of the $n$ Lorentz indices
and the subtraction of trace terms, such that contracting any pair of indices yields zero.
The coefficients $c_{f;\mu \nu \mu_1 \dots \mu_n} (q)$ are given by
\begin{equation}
\begin{split}
c_{f;\mu \nu \mu_1 \dots \mu_n} (q) = & 
2 \mathcal Q_f^2 \bigl( -g_{\mu\nu}  + \frac{q_{\mu} q_{\nu}}{q^2} \bigr) 2^n \frac{q_{\mu_1} \dots q_{\mu_n}}{(Q^2)^n} \: + \\
& 2 \mathcal Q_f^2 \bigl( g_{\mu \mu_1} - \frac{q_{\mu} q_{\mu_1}}{q^2}  \bigr) \bigl( g_{\nu \mu_2} - \frac{q_{\nu} q_{\mu_2}}{q^2}  \bigr)
2^n \frac{q_{\mu_3} \dots q_{\mu_n}}{(Q^2)^{n-1}} \: 
\end{split}
\end{equation}
and $M_{fj}(Q, \tilde\mu)$ are the coefficients of the operator mixing. 
To leading order in perturbation theory, $M_{fj}^{LO} = \delta_{fj}$, and to next-to-leading order
\begin{equation}
M_{fj}^{NLO} (Q, \tilde \mu) = 
\biggl[ \biggl( \frac{\log (Q^2/\Lambda^2)}{\log (\tilde\mu^2/\Lambda^2)}  \biggr)^{a^{(n)}/2b_0}  \biggr]_{fj}
\: .
\end{equation}
In this expression, $\Lambda$ is the energy at which the one-loop renormalized coupling diverges,
$b_0=11-\frac{2}{3}n_f$ the first coefficient of the expansion of the beta function in powers of the gauge coupling $g$
and $\gamma_{ij}^{(n)} = -\frac{g^2}{(4 \pi)^2} a_{ij}^{(n)}$ is the anomalous-dimension matrix for the twist-two operators of dimension $n+2$.
The index $j$ runs over $\{f,g\}$.

The strategy we follow to obtain moment sum rules for the structure functions from the OPE is to first relate the time-ordered product 
to the corresponding retarded commutator of currents; and secondly to use the fixed-virtuality dispersive representation of the 
retarded correlator. Details of the derivation are given in appendix~\ref{app:sum_rules}. The result is 
\begin{equation}
\int_0^{\infty} \dint x \: x^{n-1} \: [F_1 (x, Q^2)]_{\textrm{leading-twist}} =  
\sum_{f,j} \frac{\mathcal Q_f^2}{2} M_{fj} (Q, \tilde \mu) \langle O_{nj} \rangle \: , \quad n = 2,4,\dots \: ,
\label{eq:F1_sum_rule}
\end{equation}
where
\begin{equation}\la{eq:Onj_red_def}
\langle O_{nj}^{\mu_1 \dots \mu_n} \rangle = T^n [u^{\mu_1} \dots u^{\mu_n}  - \mathrm{traces}]\langle O_{nj} \rangle \: .
\end{equation}
Apart from the upper integration limit being infinity rather than unity, which makes it necessary to isolate the leading-twist part
of the structure function prior to computing its moments, the form of the moment sum rule (\ref{eq:F1_sum_rule})
is the same as in standard DIS.
In appendix \ref{sect:freeF1}, we verify these moment sum rules at leading order in the case of non-interacting quarks
by separately computing the left- and right-hand side.
Given that $\<O_{2j}\>$ is proportional to the enthalpy, that the total momentum of the fluid is generically given by the product of its enthalpy with its four-velocity,
and that $F_1$ has a parton-model interpretation as being proportional to the parton distribution with a known prefactor (Eq.\ (\ref{eq:F1f}) below),
the $n=2$ sum rule expresses in this case the (common sense) fact that the quarks carry the entire momentum of the fluid.
Next, we formulate specifically the $n=2$ sum rule for Wilson coefficients at next-to-leading order (NLO) accuracy,
which is appropriate for QCD and significantly modifies the OPE prediction for the $n=2$ moment of the structure function $F_1$.

\subsection{The $n = 2$ sum rule to NLO}

We now formulate the lowest moment sum rule at NLO accuracy in the Wilson coefficients.
We follow the notation of Ref.~\cite{Peskin:1995ev}.
The twist-two fermionic local operator renormalized at a scale $Q^2$ is, at NLO,
\begin{equation}
\begin{split}
\bigl[ O_{2f}^{\mu \nu} \bigr]_Q & = \frac{1}{2} \frac{1}{16/3 + n_f} T^{\mu \nu}_{\mathrm{traceless}} \\
& + \frac{1}{n_f(16/3 + n_f)} \biggl( \frac{\log (Q^2/\Lambda^2)}{\log (\tilde \mu^2/\Lambda^2)} 
\biggr)^{-\frac{4}{3} (\frac{16}{3} + n_f)/2b_0}
\biggl[ \frac{16}{3} \sum_{f'} O_{2f'}^{\mu \nu} - n_f O_{2g}^{\mu \nu} \biggr]_{\tilde \mu} \\
& + \biggl( \frac{\log (Q^2/\Lambda^2)}{\log (\tilde \mu^2/\Lambda^2)} \biggr)^{-32/9b_0}
\biggl[ O_{2f}^{\mu \nu} - \frac{1}{n_f} \sum_{f'} O_{2f'}^{\mu \nu} \biggr]_{\tilde \mu}\;,
\end{split}
\end{equation}
where $n_f$ is the number of fermion flavors, and $T^{\mu \nu}_{\mathrm{traceless}} = 2 \bigl( \sum_f O_{2f}^{\mu \nu} + O_{2g}^{\mu \nu} \bigr)$
is the traceless part of the energy-momentum tensor.
In the absence of a flavor-non-singlet chemical potential, the in-medium expectation value of the last term vanishes by SU($n_f$) flavor symmetry.
As a consequence, the moment sum rule to next-to-leading order in $\alpha_s$ is
\begin{equation}
\begin{split}
& \int_0^{\infty} \dint x \: x\, [F_1 (x,Q^2)]_{\textrm{leading-twist}}  = 
\sum_f \frac{\mathcal Q^2_f}{2} \biggl[  \frac{1}{2} \frac{1}{16/3 + n_f} \langle T \rangle \\ 
& \qquad \quad + \frac{1}{n_f(16/3 + n_f)} \biggl( \frac{\log (Q^2/\Lambda^2)}{\log (\tilde \mu^2/\Lambda^2)} 
\biggr)^{-\frac{4}{3} (\frac{16}{3} + n_f)/2b_0}
\bigl(\frac{16}{3} \sum_{f'} \langle O_{2f'} \rangle - n_f \langle O_{2g} \rangle \bigr) 
 \biggr] \: ,
\end{split}
\end{equation}
where
\begin{equation}
\begin{split}
& \langle O_{2j}^{\mu \nu} \rangle = T^2  \biggl[ u^{\mu} u^{\nu} - \frac{1}{4} g^{\mu \nu}\biggr] 
\langle O_{2j} \rangle \,,\\
& \langle T^{\mu \nu}_{\mathrm{traceless}} \rangle = T^2  \biggl[ u^{\mu} u^{\nu} - \frac{1}{4} g^{\mu \nu}\biggr] 
\langle T \rangle \: .
\end{split}
\end{equation}
Given the energy-momentum tensor of a relativistic fluid at thermal equilibrium
\begin{equation}
T^{\mu \nu} = (e + p) u^{\mu} u^{\nu} - p g^{\mu \nu} \: ,
\end{equation}
we find
\begin{equation}
\langle T \rangle = \frac{e + p}{T^2} \: ,
\end{equation}
where $e$ is the energy density and $p$ the pressure, their sum corresponding to the enthalpy density.
In the extreme $Q \to \infty$ limit, we have
\begin{equation}\la{eq:n=2momentNLO}
\lim_{Q^2\to\infty}\int_0^{\infty} \dint x \: x \: [F_1 (x, Q^2)]_{\textrm{leading-twist}} =
\bigl( \sum_f {\mathcal Q_f^2} \bigr) \frac{1}{16/3 + n_f} \frac{e + p}{4T^2}
\: .
\end{equation}
Thus the asymptotic $n=2$ moment of the structure function $F_1$ is
uniquely determined by the enthalpy density of the medium. By invoking
the interpretation of $F_1$ in terms of a parton distribution function
(see Eq.\ (\ref{eq:F1f}) below), and recalling that the total momentum
density carried by the fluid is $(e+p)\vec u$,
Eq.\ (\ref{eq:n=2momentNLO}) also shows that the asymptotic momentum
fraction carried by one out of $n_f$ flavors of quarks is $1/(n_f +
16/3)$, while the gluons carry the fraction $(16/3)/(n_f+
16/3)$. These fractions are exactly the same as in conventional DIS on
the nucleon. In that context, these momentum fractions only apply at
extremely large $Q^2$.  For $n_f=3$, the asymptotic gluon momentum
fraction is 0.64. This number is substantially larger than the
enthalpy fraction contributed by the gluons in the high-temperature
limit, which is about 0.34. This difference reflects the fact that
the non-interacting structure function $F_1$ satisfies the $n=2$ sum rule
with leading-order Wilson coefficients, and not the NLO sum rule (\ref{eq:n=2momentNLO})
appropriate for QCD. Thus the $n=2$ moment of the structure function $F_1$ in the DIS limit is unstable against
`turning on' interactions between quarks, and cannot be expanded in powers of $g^2(T)$.
We return to this point in the conclusion.

\section{Tests of the onset of Bjorken scaling from Euclidean correlation functions\la{sec:fixedQ2}}

An interesting question is how large the photon virtuality has to be
in order for the structure functions at fixed $x$ to become
$Q^2$-independent. It is tempting to address this question by
computing both sides of a moment sum rule, however the $x$-moments of
the spectral functions correspond to derivatives of Euclidean
correlators at vanishing Matsubara frequency, and these derivatives are
only accessible via a numerically ill-posed analytic continuation.  We
therefore propose tests that do not require a numerical analytic
continuation.

In the rest frame of the plasma,
the spectral function components are related to the longitudinal and transverse spectral functions $\sigma^L$ and $\sigma^T$,
as defined in Ref.~\cite{Meyer:2018xpt}, as well as to the spectral functions $\rho_1$ and $\rho_2$ introduced in the DIS context according to
\begin{equation}
\begin{split}
\frac{1}{2} \bigl(\delta^{ij} - \frac{q^i q^j}{\vec q^2} \bigr)\rho^{ij} (u, q) =  -\sigma^T(q^0, Q^2)  &= \rho_1 (q^0, Q^2) \;, \\
\rho^{00} (u, q) = \frac{\vec q^2}{Q^2} \sigma^L(q^0,Q^2) &= -\frac{\vec q^2}{Q^2} \rho_1 (q^0, Q^2) + \frac{T |\vec q|^4}{q^0 Q^4} \rho_2(q^0, Q^2)\;, \\
\frac{q^i q^j}{\vec q^2}\rho^{ij} (u, q) = \frac{(q^0)^2}{Q^2} \sigma^L(q^0,Q^2)  &=-\frac{(q^0)^2}{Q^2} \rho_1 (q^0, Q^2) + \frac{T q^0 \vec q^2}{Q^4} \rho_2(q^0, Q^2)
\: .
\end{split}
\label{eq:spf_components}
\end{equation}
The two longitudinal and transverse components of the retarded polarisation tensor are denoted as
 $G_R^L(q^0,\vec q^2)$ and $G_R^T(q^0,\vec q^2)$; see Eq.\ (\ref{eq:GR}) for their explicit definition.
The imaginary part, for $q^0$ approaching the real axis from above, yields the corresponding spectral function.
At a spacelike or lightlike point $Q^2\geq 0$, the relation to the spectral functions $\sigma^{L,T}$ reads
\be
2\,{\rm Im}\,G_R^{L,T}(q^0,\vec q^2) =
\sigma^{L,T}(q^0,Q^2), \qquad Q^2= \vec q^2 - (q^0)^2.
\ee

For imaginary $q^0 = i \omega_n$, with $\omega_n = 2 n \pi T>0$ the $n$-th Matsubara frequency, the retarded correlator
coincides with the Euclidean correlator~\cite{Meyer:2011gj},
\begin{equation}
G_{R}^{L,T} \bigl(i \omega_n, \vec q^2 \bigr) = 
G_{E}^{L,T} \bigl(\omega_n, \vec q^2   \bigr) \equiv 
H_E^{L,T} (\omega_n; Q^2) \: , \qquad Q^2 = \vec q^2 + \omega_n^2 .
\label{eq:Eucl_corr}
\end{equation}
For a given spacelike virtuality $Q^2$, the Euclidean correlators can be obtained in lattice QCD as a function of Matsubara frequency by computing
\ba\la{eq:H_E_T}
H_E^{T} (\omega_n; Q^2) &=& 
\int d^4x\; e^{\sqrt{\omega_n^2-Q^2}\hat{\vec q}\cdot \vec x + i\omega_n x_0}\;
\Big({\txts\frac{1}{2}} \bigl(\delta_{ik} - \hat q_i \hat q_k \bigr)\; \< j_i(x)\,j_k(0)\>\Big),
\\
H_E^{L} (\omega_n; Q^2) &=& \int d^4x\; e^{\sqrt{\omega_n^2-Q^2}\hat{\vec q}\cdot \vec x + i\omega_n x_0}\;
\Big(\<j_0(x)\,j_0(0)\> + \hat q_i \hat q_k  \< j_i(x)\,j_k(0)\>\Big).
\la{eq:H_E_L}
\ea
On the right-hand side of these equations, $\hat{\vec q}$ is the unit vector representing the direction
of the spatial momentum, and we are using Euclidean notation, implying a sign change in the two-point
functions of the spatial current relative to its Minkowski-space correspondent.
We have anticipated the fact that in order to probe DIS kinematics ($Q^2 ={\rm O}(Tq^0)$),
the spatial momentum in the Euclidean correlator needs to become imaginary and close in magnitude
to $\omega_n$. Thus the weight function of the coordinate-space correlator is a real exponential, as
for the case of vanishing virtuality~\cite{Meyer:2018xpt}.
The dispersion relation (\ref{eq:disp_rel_omega}) below however also applies
to the case where one or more of the spatial momenta are real, as long a the virtuality is kept fixed.
We also remark that Eqs.\ (\ref{eq:H_E_T}--\ref{eq:H_E_L}) could be averaged analytically
over the direction of $\hat{\vec q}$ using Legendre polynomials, along the lines of~\cite{Meyer:2017hjv}.
This could prove numerically advantageous in calculations performed on lattices with a large spatial volume.

The fixed-virtuality dispersion relation \cite{Meyer:2018xpt}
\begin{equation}
H_E^{L,T} (\omega_n; Q^2) - H_E^{L,T} (\omega_r; Q^2) = 
\int_0^{\infty} \frac{ \dint \omega}{\pi} \: \omega \: \sigma^{L,T} (\omega, Q^2) 
\biggl[ \frac{1}{\omega^2 + \omega_n^2} - \frac{1}{\omega^2 + \omega_r^2} \biggr] \: 
\label{eq:disp_rel_omega}
\end{equation}
follows.
The subtraction of $H_E^{L,T} (\omega_r;Q^2)$ is needed to ensure the convergence of the dispersive integral\footnote{The forward Compton scattering
amplitudes of the proton require analogous subtractions. We also remark that, in the lattice regularization of QCD,
an additional subtraction is in general needed on the right-hand side of Eq.~(\ref{eq:disp_rel_omega})
due to an ultraviolet divergence; see \cite{Meyer:2018xpt} for one  such suitable subtraction.}.
We can rewrite Eq.~\ref{eq:disp_rel_omega} as an integral over the variable $x = Q^2/(2 T \omega)$ as follows 
\begin{equation}
H_E^{L,T} (\omega_n; Q^2) - H_E^{L,T} (\omega_r; Q^2) = 
\int_0^{\infty} \frac{ \dint x}{\pi} \: x \: \hat\sigma^{L,T} (x, Q^2) 
\frac{a_r^2 - a_n^2}{(1 + a_n^2 x^2)(1 + a_r^2 x^2)}
\: , 
\label{eq:disp_rel_x}
\end{equation}
where $a_n \equiv 2 T \omega_n / Q^2$ and for convenience we have defined spectral functions that take as arguments the Bjorken variables,
$\hat\sigma^{L,T} (x, Q^2) = \sigma^{L,T} (Q^2/(2 T x), Q^2)$.
The Euclidean correlators $H_E^{L}$ and $H_E^T$ can be computed directly in lattice QCD.
Thus Eq.\ (\ref{eq:disp_rel_x}) can be used  to probe the $Q^2$
evolution of the spectral functions $\hat\sigma^{L,T}$. For this purpose, the kinematic variables $a_n$ and $a_r$
should be kept fixed as $Q^2\to\infty$. In particular, the extent to which the
spectral functions become $Q^2$-independent at fixed $x$ can be probed
via the corresponding Euclidean correlators, albeit a global test involving a weighted integral over $x$.
The typical $x$-values contributing are of order unity, provided $a_n$ and $a_r$ are chosen of that order.

The spectral function $\hat\sigma^{T}$ coincides up to its sign with $\rho_1$, while
using $x=Q^2/(2Tq^0)$, it follows from Eqs.\ (\ref{eq:spf_components}) that
\begin{equation}
\begin{split}
\frac{q^i q^j}{\vec q^2}\rho^{ij} (u, q) - \rho^{00} (u, q) = -\sigma^L(q^0,Q^2)
& = \left(\rho_1(q^0, Q^2) - \frac{\rho_2(q^0, Q^2)}{2x}\right)   -   \frac{2 x T^2}{Q^2}  \rho_2(q^0, Q^2) 
\: .
\end{split}
\label{eq:polarizations}
\end{equation}
Thus the longitudinal spectral function $\sigma^L$, which can be probed in lattice QCD via Eq. (\ref{eq:disp_rel_x}), 
measures the size of the correction to the Callan-Gross relation (\ref{eq:CGrho}),
up to a term suppressed by the factor $1/Q^{2}$ in the Bjorken limit. 
Since in QCD the Callan-Gross relation is asymptotically violated by terms of order $\alpha_s(Q^2)$~\cite{Manohar:1992tz}, 
where $\alpha_s =g^2(Q^2)/(4\pi)$ is the running coupling, the power-suppressed term is parametrically irrelevant.
Thus, in the Bjorken limit, one expects $H_E^{L}$ to be suppressed (logarithmically in $Q^2$) as compared to $H_E^{T}$.
It is interesting at this point to note that the longitudinal channel $H_E^{L}$ vanishes for $Q^2=0$, while 
$H_E^{T}$ probes the real-photon emissivity. Thus, while the Bjorken limit is very far from lightlike kinematics in the sense that 
$Q^2\gg T^2$, a similar suppression of the longitudinal relative to the transverse channel is expected.

\begin{figure}
\includegraphics[width = .5\textwidth]{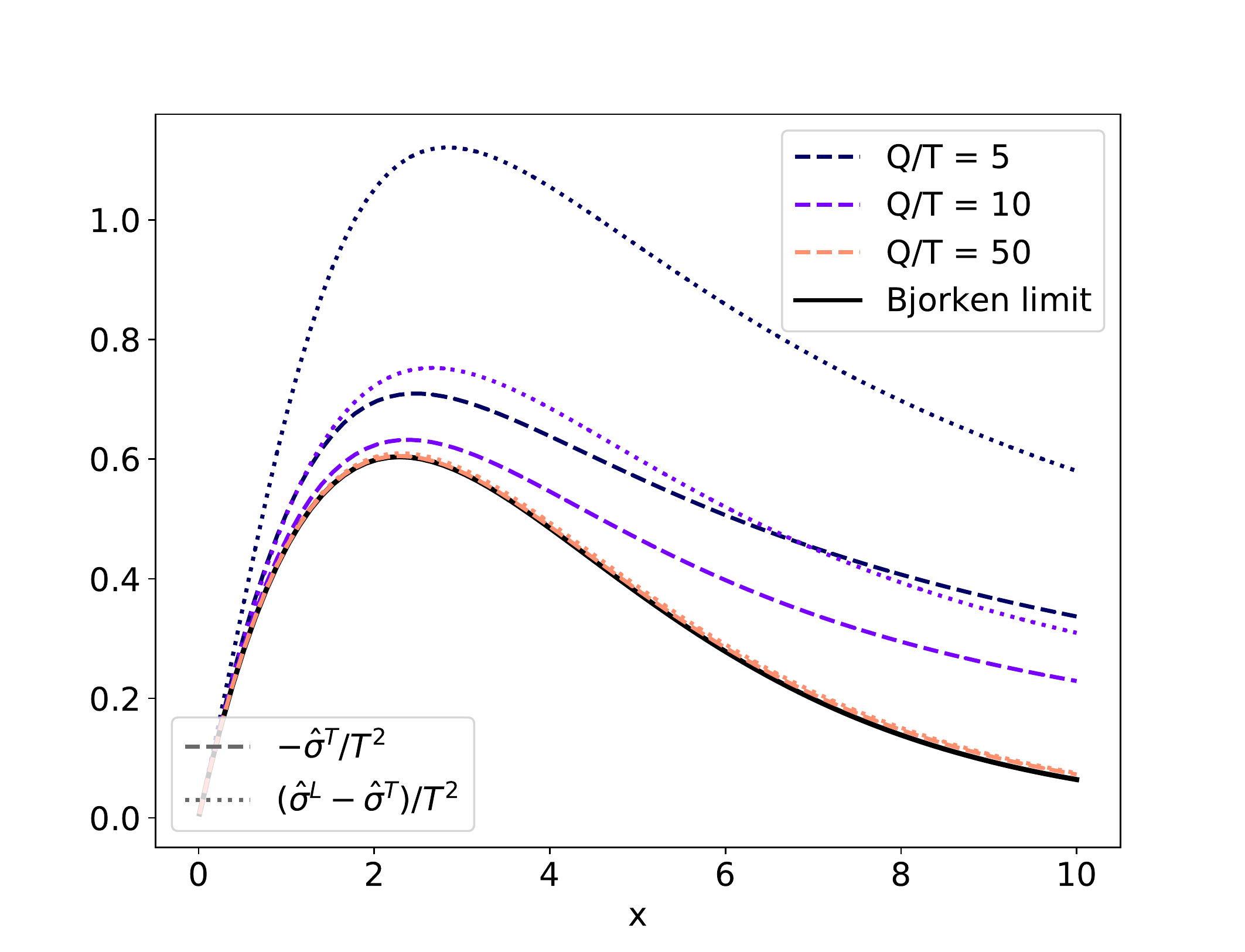}~
\includegraphics[width = .5\textwidth]{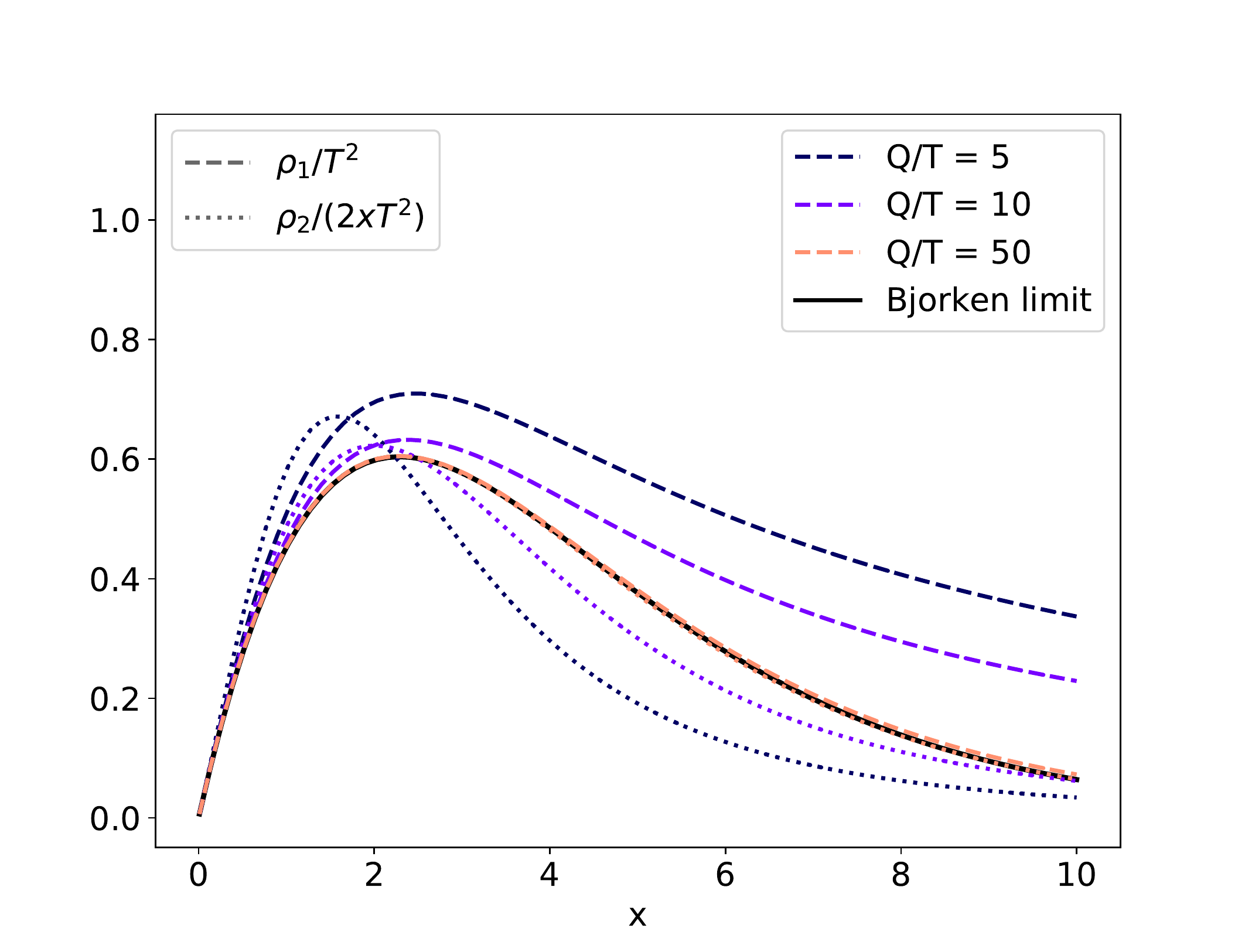}
\caption{Left: Approach to the Bjorken limit of the transverse as well as transverse-minus-longitudinal
  spectral function, in the case of a plasma of free fermions. The free spectral functions from~\cite{Laine:2013vma}
  have been used.
Right: Breaking of the Callan-Gross relation $\rho_2 = 2x\rho_1$ at finite $Q^2$ in the free theory.
The functions $\rho_1$ and $\rho_2$ have been computed from the free spectral functions using Eq.~(\ref{eq:spf_components}).
In both panels, $N_c=3$ and $\sum_f {\cal Q}_f^2=1$ has been chosen.
}
\label{fig:Callan-Gross_free}
\end{figure}

In Fig.~\ref{fig:Callan-Gross_free}, left panel, we illustrate the
approach of $-\hat\sigma^{T}$ and $(\hat\sigma^{L} - \hat\sigma^{T} )$ to
their common Bjorken limit in the free theory. We observe that in both
cases the departure from the limit value increases with $x$, and that
$-\hat\sigma^{T}$ approaches this limit faster. The right panel allows one
to judge the size of the breaking of the Callan-Gross relation as a function of $x$
in relation to the magnitude of $\rho_1$.

\begin{figure}
\includegraphics[width = .8\textwidth]{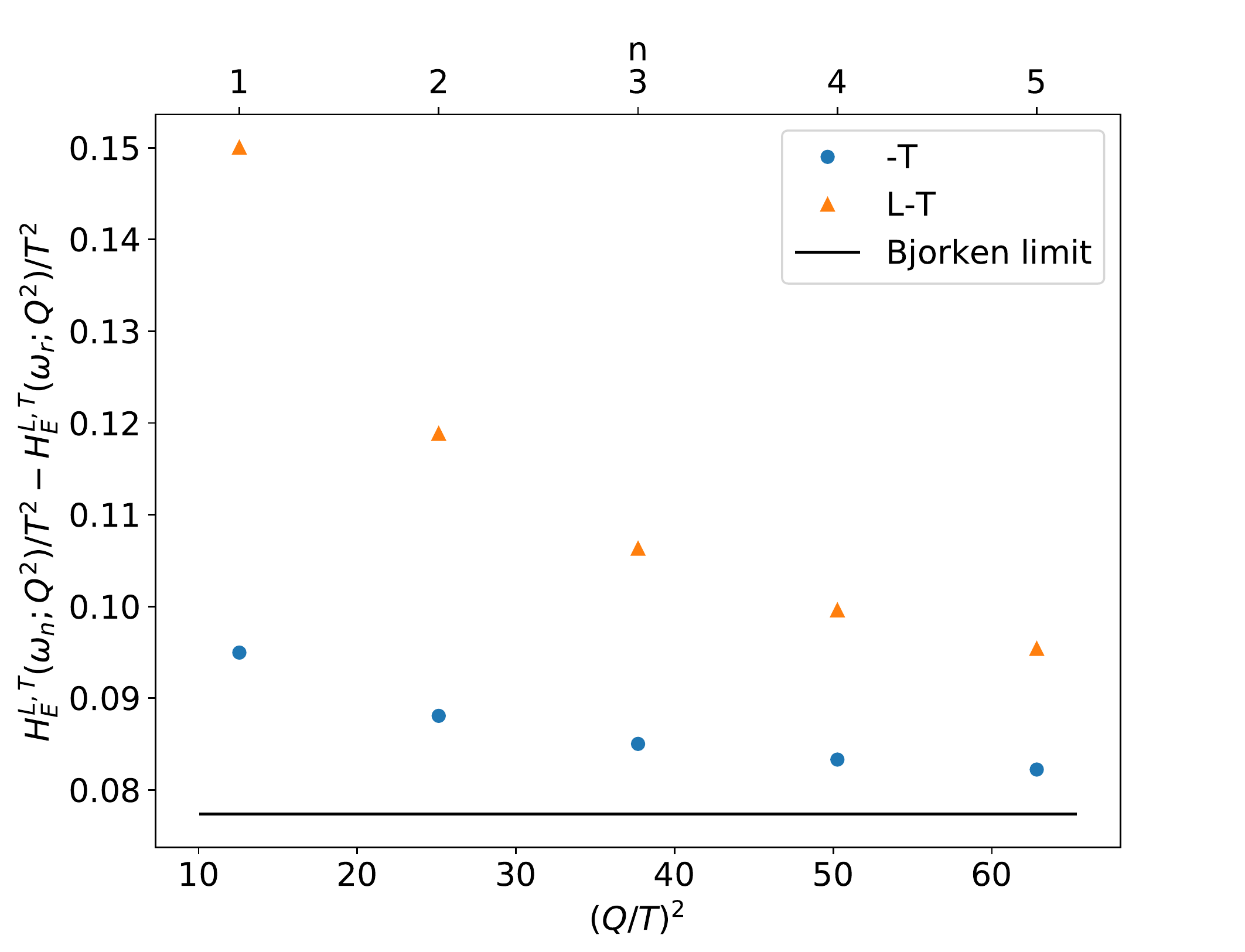}
\caption{Fixed-virtuality correlation functions (\ref{eq:disp_rel_x}) in the free theory, as functions of $Q^2$.
As we increase $\omega_n$, we set $Q^2 = 2 T \omega_n$ and $\omega_r = 2 \omega_n$, i.e.\ $a_n = 1$ and $a_r = 2$ are kept constant.
Two choices of polarization are shown, that converge to the same value in the Bjorken limit.
The difference between these two polarizations measures the breaking of the Callan-Gross relation at the
correlator level, up to a $1/Q^2$ suppressed term (see Eq.~(\ref{eq:polarizations})).
}
\label{fig:HE_free}
\end{figure}

As noted below Eq.\ (\ref{eq:disp_rel_x}), by computing the correlators on the left-hand side of the equation
on the lattice for increasing values of the virtuality $Q^2$,
one could observe the approach to Bjorken scaling.                
In Fig.~\ref{fig:HE_free}, we show as an example the corresponding
analysis performed in the free theory.  By using the spectral
functions given in Ref.~\cite{Laine:2013vma}, we integrate numerically
the right-hand-side of Eq.~\ref{eq:disp_rel_x}.  In order to keep
$a_n$ and $a_r$ fixed and $O(1)$ as we increase $Q^2$, we set
$\omega_n = \omega_r/2 = Q^2/(2T)$, a choice that corresponds to $a_n= 1$ and $a_r = 2$.
Figure \ref{fig:HE_free} displays the difference of Euclidean
correlators for two choices of polarization which, due to the
Callan-Gross relation, converge to the same value in the Bjorken
limit. The figure illustrates again that the convergence is faster
for the transverse channel $H_E^T$ than for the difference $(H_E^L - H_E^T)$.
One can also read off from the figure how suppressed $H_E^L$ is in comparison to $H_E^T$.

\section{Parton model\la{sec:parton}}
In this section, we review the partonic interpretation of the structure functions of the thermal medium.
The steps follow closely the textbook treatment of unpolarized DIS on the nucleon, 
a small but significant difference being the parametrization of the parton momentum; see Eq.\ (\ref{eq:pxi}) below.

Consider the elastic electron--parton collision via one-photon exchange; see the right panel of figure \ref{fig:kin}.
The parton is assumed to be a Dirac particle and to carry an electric charge equal to ${\cal Q}_f$ times the electron charge.
It has initial momentum $p_\xi$, while the electron has initial momentum $k$.
The differential cross-section reads  (see e.g.\ Ref.~\cite{Peskin:1995ev}, Eq.~(14.3))
\be
\frac{d\sigma}{d\hat t} = \frac{2\pi\alpha^2}{\hat s^2} \,{\cal Q}_f^2 \; \frac{\hat s^2 + (\hat s+\hat t)^2}{\hat t^2},
\ee
where $\hat s$ and $\hat t$ are the Mandelstam variables of the electron--parton system.
The variable $\hat t=q^2 = -Q^2$ is the virtuality of the photon and coincides with the Mandelstam variable $t$ of the 
electron--fluid-cell collision. On the other hand, 
$\hat s = (p_\xi +k)^2 \approx 2p_\xi\cdot k$
is specific to the electron--parton collision.

In a frame in which the fluid cell has a large $\gamma$ factor,
we express the fact that $p_\xi$ is to a good approximation collinear with the fluid velocity $u$,
\be\la{eq:pxi}
p_\xi = \xi \, m_T \, u.
\ee
Let $f_f(\xi)\, d\xi$ represent the number of partons of type $f$ in the fluid cell carrying momentum $\xi \: m_T \: u$.
One does not distinguish between quarks and antiquarks, e.g.\ for the up quark
$f_u(\xi) = u_\uparrow(\xi)+u_\downarrow(\xi)+\bar u_\uparrow(\xi)+\bar u_\downarrow(\xi)$. One expects
this number to be of order the volume of the cell.
Translating the electron--parton cross-section into an electron--fluid-cell cross-section, we obtain (see~\cite{Peskin:1995ev}  Eq.~(14.8))
\be\la{eq:dsig_parton}
\frac{d\sigma}{dQ^2} = 
\sum_f {\cal Q}_f^2 \,f_f(\xi) d\xi\; \frac{2\pi\alpha^2}{Q^4}\,\Big[1+\Big(1-\frac{Q^2}{2 \xi m_T (u \cdot k)}\Big)^2\Big].
\ee
Recalling the definition (Eq.\ (\ref{eq:x.and.y})) of the two dimensionless kinematic variables $x$ and $y$,
 Eq.\ (\ref{eq:dsig_parton}) becomes
\be\la{eq:dsig_parton2}
\frac{d\sigma}{dQ^2} = \sum_f {\cal Q}_f^2 \,f_f(\xi) d\xi\; \frac{2\pi\alpha^2}{Q^4}\,
\Big[1+\Big(1-\frac{xy}{\xi}\Big)^2\Big].
\ee

We found Eq.\ (\ref{eq:dsigdEpdOm})  for $\frac{d^2\sigma}{dE'd\Omega}$ in the fluid rest frame 
in terms of its spectral functions.
Using the conversion\footnote{First, show that $\frac{d^2\sigma}{dE'd\Omega}= \frac{d^2\sigma}{dxdyd\phi}\cdot \frac{E'}{m_T (u\cdot q)}$
by showing that $\abs{\frac{\partial(x,y)}{\partial(E',\cos\theta)}}= \frac{E'}{m_T(u\cdot q)}$;
for that, the relations $x = \frac{2 E E' (1 - \cos\theta)}{2 m_T (E - E')}$ and $y= \frac{E - E'}{E}$ are useful.
Second, show that $\frac{d^2\sigma}{dxdyd\phi} = \frac{d^2\sigma}{dxdQ^2d\phi}\, \frac{Q^2}{y}$ by showing 
that  $\frac{\partial(x,Q^2)}{\partial(x,y)}= \frac{Q^2}{y}$. For that, the relation 
$Q^2 = 2 x y E m_T$ is useful.} (see \cite{Manohar:1992tz}, Eq. (2.7))
\be
\frac{m_T (u\cdot q)}{E'} \; \frac{d^2\sigma}{dE'd\Omega} = \frac{Q^2}{y}\, \frac{d^2\sigma}{dxdQ^2d\phi}\;,
\ee
the cross-section (\ref{eq:dsigdEpdOm}) becomes
\be\la{eq:dsig_strucfct}
\frac{d^2\sigma}{L^3dQ^2dx} = \frac{2 m_T}{x}  \frac{\alpha^2}{Q^4(1-e^{-\beta q^0})}\,
\Big[xy^2\rho_1(u \cdot q,Q^2)+(1-y)\rho_2(u \cdot q,Q^2)\Big].
\ee
Note that Eq.\ (\ref{eq:dsig_strucfct}) corresponds to the cross-section for scattering
on the full thermal system, which occupies a volume $L^3$ in its rest-frame.
In this section, we are considering the lepton scattering on a particular subvolume of the full system.
Since the cross-section scales with the volume, we now reinterpret the factor $L^3$ as being the
rest-frame volume of that particular fluid cell.

The cross-section (\ref{eq:dsig_strucfct}) is in a form to which the parton-model calculation can easily be matched.
By analogy with DIS on the proton, we anticipate that $\xi$ is proportional to $x$.
Noting that the cross-section has the form of a quadratic polynomial in $y$,
the terms in the square bracket proportional to $-y$ and $1$ are equal in Eq.\ (\ref{eq:dsig_strucfct}), 
and they are also equal in Eq.\ (\ref{eq:dsig_parton2}) precisely when 
\be
x/\xi= 1 \,.
\ee
Continuing from here, matching the ratio of the term proportional to $y^2$ to the term proportional to $1$ 
between Eqs.\ (\ref{eq:dsig_strucfct}) and (\ref{eq:dsig_parton2}) yields
\be\la{eq:CGrho}
\rho_2(u \cdot q,Q^2) = 2x\rho_1(u \cdot q,Q^2).
\ee
This is the Callan-Gross relation.
Matching the absolute normalization of the cross-sections for $y=1$, one obtains
\be
\sum_f {\cal Q}_f^2 f_f(\xi) =  L^3 \frac{m_T}{\pi} \frac{\rho_1(u\cdot q, Q^2)}{1-e^{-\beta q^0}}.
\ee
Recalling the KMS relation (\ref{eq:rho12}) between spectral and structure functions, the choice $m_T = T$ allows us to rewrite
\be\la{eq:F1f}
F_1(u\cdot q,Q^2) = \frac{1}{4L^3T} \sum_f {\cal Q}_f^2 \;f_f(x).
\ee
In words, $4\,F_1\cdot dx$ is the ${\cal Q}_f^2$-weighted number of partons carrying a momentum $xT$ times the fluid four-velocity $u^\mu$
per unit transverse area in a slab of fluid which in its rest frame has thickness $1/T$ in the longitudinal direction.
This makes the choice $m_T=T$ natural.

The interpretation of the structure function as a parton distribution
function is confirmed by the following observation.  The first moment
$\int_0^\infty dx f(x)\; xT u^k$ should yield the spatial momentum of
the fluid cell carried by quarks and antiquarks.  The quantity playing
the role of momentum density in ideal hydrodynamics is $(e+p) u^k$,
therefore we would expect to find $(e+p)L^3 u^k$ for the first moment
in the case of free quarks.  The connection (\ref{eq:F1f}) of the
parton distribution function $f$ to the structure function $F_1$ together with
the $n=2$ moment sum rule obeyed by $F_1$
(Eq.\ (\ref{eq:F1_sum_rule})), which is derived from the
operator-product expansion, independently confirm this expectation;
see the remark below Eq.\ (\ref{eq:eq:F1_sum_ruleRHS2}).

\section{Conclusions and outlook\la{sec:concl}}

In this paper we have provided a physics interpretation of the two
thermal spectral functions characterizing the electric-current
correlator in the spacelike regime. They are related via the KMS
relation to the structure functions which describe the total
cross-section for a light lepton scattering on the thermal medium.
What makes this observation interesting is that a lot is known about
the physics of DIS, in particular, we have shown that the same moment
sum rules apply to the thermal QCD medium as to the nucleon.
Therefore, the first few moments of the structure functions in the DIS
limit can be computed in lattice QCD via the expectation value of
local twist-two operators.  Furthermore, Euclidean correlation
functions can be computed in lattice QCD probing the transverse and
the longitudinal spectral function at a fixed spacelike virtuality
$Q^2$, thus allowing the onset of the DIS regime to be probed.  This
approach offers a rare opportunity to answer the question, ``On what
length scales do the quarks and gluons in the quark-gluon plasma start
to become `visible'?'' in a theoretically clean way.  Importantly,
none of these lattice-QCD based studies requires addressing the
numerically ill-posed problem of resolving the $x$-dependence of the
structure functions from the Euclidean correlators.  On the other
hand, the DIS kinematic regime is numerically challenging to reach,
since it involves using an imaginary spatial momentum close in
magnitude to the Matsubara frequency, requiring one to achieve
excellent control over non-static screening correlators at long
distances. For that purpose, a dedicated study of the corresponding
non-static correlation lengths~\cite{Brandt:2014uda} is a good preparatory step.

It is well-known from the OPE analysis of standard DIS on the nucleon
that the non-trivial one-loop anomalous dimensions of the twist-two
operators and their mixing lead to a modified prediction of the
infinite-$Q^2$ limit of the structure functions, as compared to a
leading-order analysis. The same observation applies to the structure
functions of the thermal medium. As we have verified explicitly for
the first two non-trivial moments, the moment sum rules (\ref{eq:F1_sum_rule}) using
leading-order Wilson coefficients are consistent with the structure
functions of thermal, non-interacting quarks. There is no analogue of
this calculation in standard DIS, because a nucleon would simply not
exist in the absence of the ${\rm SU}(3)_{\rm color}$ gauge field,
whereas a plasma of non-interacting quarks makes perfect sense.  In
QCD however, even at a very high temperature where one might have
expected interactions to play a subleading role, the infinite-$Q^2$
limit of the structure functions differs from that of non-interacting
quarks by an amount that is not suppressed by the strong coupling
constant.
Indeed, the $Q^2\to\infty$ asymptotic momentum fractions carried by the quarks and the gluons according
to the OPE with NLO Wilson coefficients is the same as in DIS on the nucleon --
see the discussion below Eq.\ (\ref{eq:n=2momentNLO}), 
whereas explicitly taking the $n=2$ moment of the leading-order spectral function (\ref{eq:freeF1})
leads to a different result.
Thus it is clear that the structure functions of free quarks in the DIS regime are unstable against `turning on' their interactions.
Clearly, this issue deserves further investigation, which could be carried out with the help of the next-to-leading
results~\cite{Jackson:2019mop,Jackson:2019yao} for the vector spectral functions.
We also conclude from these remarks that the content of the moment sum rules is far from trivial,
and could be used to test future weak-coupling calculations of the thermal spectral functions,
including the sophisticated resummations performed in this context; see references~\cite{Arnold:2001ba,Arnold:2001ms,Ghiglieri:2013gia}
for the case of vanishing virtuality.
In particular, calculations of thermal spectral functions in the DIS regime can benefit
from the existing three-loop results for the splitting functions
and anomalous dimensions of the twist-two operators~\cite{Vogt:2004mw}.

\acknowledgments
We thank Csaba T\"or\"ok for discussions.
This work was supported by the
European Research Council (ERC) under the European Union’s Horizon
2020 research and innovation program through Grant Agreement
No.\ 771971-SIMDAMA. 
The work of M.C.\ is supported by the European Union's Horizon
2020 research and innovation program
under the Marie Sk\l{}odowska-Curie Grant Agreement No.\ 843134-multiQCD.

\appendix

\section{Derivation of the moment sum rules}
\label{app:sum_rules}
In this appendix we provide a derivation of Eq.\ (\ref{eq:F1_sum_rule}).

\subsection{OPE prediction for the invariant amplitudes of the time-ordered correlator}
Let $G_T^{\mu \nu}(q)$ be the thermal expectation value of the Fourier-transformed, time-ordered product of electromagnetic currents,
\begin{equation}
G_T^{\mu \nu}(q) = i \int \dint^4 x \: e^{i q \cdot x} \langle {\rm T} \{ j^{\mu} (x) j^{\nu} (0) \} \rangle =
\langle t^{\mu \nu} (q) \rangle.
\end{equation}
In the limit $Q^2 = -q^2 \to \infty$, the OPE of $t^{\mu \nu} (q)$ given in Eq.~(\ref{eq:OPE}) holds. 
Recalling the definition (\ref{eq:Onj_red_def}) of the reduced thermal expectation values $\langle O_{nj} \rangle$
in terms of their corresponding twist-two operator,
\begin{equation}
\langle O_{nj}^{\mu_1 \dots \mu_n} \rangle = T^n [u^{\mu_1} \dots u^{\mu_n} - \mathrm{traces} ] \langle O_{nj} \rangle \: ,
\end{equation}
we have
\begin{equation}
\begin{split}
G_T^{\mu \nu} (q) & \overset{Q \to \infty}{\sim} \sum_{n = 2,4,\dots} \sum_{f,j} 2 \mathcal Q_f^2 M_{fj} (Q,\tilde \mu)
\biggl[ \bigl( -g^{\mu \nu} + \frac{q^{\mu} q^{\nu}}{q^2} \bigr) \bigl( \frac{2 T (u \cdot q)}{Q^2} \bigr)^n \langle O_{nj} \rangle \\
& + \bigl( u^{\mu} - (u \cdot q) \frac{q^{\mu}}{q^2} \bigr)\bigl( u^{\nu} - (u \cdot q) \frac{q^{\nu}}{q^2} \bigr)
\bigl( 2 T \bigr)^n \frac{(u \cdot q)^{n-2}}{(Q^2)^{n-1}} \langle O_{nj} \rangle \biggr] \: .
\end{split}
\end{equation}
We consider the  decomposition of the time-ordered correlation function $G_T^{\mu \nu}$ in invariant amplitudes,
\begin{equation}
G_T^{\mu \nu} (q) = \tilde F_1 (u \cdot q, Q^2) \bigl( -g^{\mu \nu} + \frac{q^{\mu} q^{\nu}}{q^2}  \bigr) 
+ \tilde F_2 (u \cdot q, Q^2) \frac{T}{u \cdot q} \bigl( u^{\mu} - (u \cdot q) \frac{q^{\mu}}{q^2}  \bigr)
\bigl( u^{\nu} - (u \cdot q) \frac{q^{\nu}}{q^2}  \bigr)
\: ,
\label{eq:strf}
\end{equation}
and find the following expansions for the invariant amplitudes,
\begin{eqnarray}
\tilde F_1 (u \cdot q, Q^2) & \overset{Q \to \infty}{\sim} &
\sum_{n = 2,4,\dots} \sum_{f,j} 2 \: \mathcal Q_f^2 \: M_{fj} (Q, \tilde \mu) \: \bigl( \frac{2 T (u \cdot q)}{Q^2} \bigr)^n
\langle O_{nj} \rangle
\label{eq:F1exp}
\\
\tilde F_2 (u \cdot q, Q^2) & \overset{Q \to \infty}{\sim} &
\sum_{n = 2,4,\dots} \sum_{f,j} 4 \: \mathcal Q_f^2 \: M_{fj} (Q, \tilde \mu) \: \bigl( \frac{2 T (u \cdot q)}{Q^2} \bigr)^{n-1}
\langle O_{nj} \rangle \: .
\end{eqnarray}

\subsection{Dispersive integral for the retarded correlator}
\label{app:disp_rel}

In the rest frame of the plasma, we introduce the retarded and advanced correlators, as well as the spectral function,
\begin{equation}
G_R^{\mu \nu} (q) = i \int \dint^4 x \:  e^{i q \cdot x} \: \theta (x^0) \: \langle [ j^{\mu}(x), j^{\nu} (0) ] \rangle
\label{eq:GR}
\end{equation}
\begin{equation}
G_A^{\mu \nu} (q) = - i \int \dint^4 x \:  e^{i q \cdot x} \: \theta (-x^0) \: \langle [ j^{\mu}(x), j^{\nu} (0) ] \rangle
\label{eq:GA}
\end{equation}
\begin{equation}\la{eq:rho_def}
\rho^{\mu \nu} (q) = \int \dint^4 x \: e^{i q \cdot x} \: \langle [ j^{\mu}(x), j^{\nu} (0) ] \rangle =
-i  (G_R^{\mu \nu} (q) - G_A^{\mu \nu} (q))
\end{equation}
We also introduce functions that exhibit the virtuality dependence explicitly,
\begin{equation}
H_{R,A}^{\mu \nu} (q^0, Q^2;\;\hat q) \equiv  
G_{R,A}^{\mu \nu} \bigl(q^0, \vec q = \sqrt{(q^0)^2 + Q^2} \: \hat q \bigr) \: ,
\end{equation}
where $\hat q \equiv \vec q / \abs{\vec q}$ is a unit vector.
We now assume that, at fixed $Q^2$, $H_R^{\mu\nu}$ is analytic for $\Im q^0 > 0$ and $H_A$ for $\Im q^0 < 0$,
as was shown in~\cite{Meyer:2018xpt} in the lightlike case $Q^2=0$. Note that the same property holds
in the case of DIS on the nucleon~\cite{Peskin:1995ev}. We define
\begin{equation}
H^{\mu \nu} (q^0, Q^2;\;\hat q) \equiv
\begin{cases}
H_R^{\mu \nu} (q^0, Q^2;\;\hat q) & \Im q^0 > 0 \\
H_A^{\mu \nu} (q^0, Q^2;\;\hat q) & \Im q^0 < 0
\end{cases} \: .
\end{equation}
We now integrate along the contour $C$ in the complex-$q^0$ plane consisting of a large half-circle in the upper plane with its diameter
running along the real axis from left to right, and a second half-circle in the lower plane with its diameter running along the real axis from right to left;
see Fig.~1 in Ref.~\cite{Meyer:2018xpt}. We obtain
\begin{equation}
\begin{split}
\frac{1}{2 \pi i} \oint_{C} \frac{H^{\mu \nu} (\tilde q^0, Q^2;\;\hat q)}{(\tilde q^0 - q^0)^m} \: \dint \tilde q^0 & =
\frac{1}{(m - 1)!} \biggl[ \frac{\dint^{m-1}}{\dint (\tilde q^0)^{m-1}} \: H^{\mu \nu} (\tilde q^0, Q^2;\;\hat q) \biggr]_{\tilde q^0 = q^0} \\
& =  \frac{1}{2 \pi} \int_{-\infty}^{\infty} \frac{\rho^{\mu \nu} (\tilde q^0, Q^2;\;\hat q)}{(\tilde q^0 - q^0)^m} \: \dint \tilde q^0 \: ,
\quad m = 1,2,\dots \: .
\end{split}
\end{equation}
Here $\rho^{\mu \nu} (q^0, Q^2;\;\hat q) \equiv \rho^{\mu \nu}(q^0, \vec q = \sqrt{(q^0)^2 + Q^2} \: \hat q)$ is the fixed-virtuality spectral function.
Setting $q^0 = i \epsilon$, we obtain
\begin{equation}
\frac{1}{(m - 1)!} \biggl[ \frac{\dint^{m-1}}{\dint (\tilde q^0)^{m-1}} \: H_R^{\mu \nu} (\tilde q^0, Q^2;\;\hat q) \biggr]_{\tilde q^0 = i\epsilon} 
 =  \frac{1}{2 \pi} \int_{-\infty}^{\infty} \frac{\rho^{\mu \nu} (\tilde q^0, Q^2;\;\hat q)}{(\tilde q^0-i\epsilon)^m} \: \dint \tilde q^0 \: .
\label{eq:disp_integral}
\end{equation}

\subsection{Relation between the retarded and the time-ordered correlator}
\label{app:T-R-hadr}

The relation of $W_>^{\mu\nu}$ to the time-ordered correlator can be worked out starting from the identity
\be
 {\rm T} \left\{j^\mu(x)\,j^\nu(0)\right\}
=  \theta(x^0) \,[j^\mu(x),\, j^\nu(0)] +  j^\nu(0)j^\mu(x).
\ee
Second, using translational invariance, we have 
\be
\int d^4x\; e^{iq\cdot x}\, \<n|j^\nu(0)\,j^\mu(x)|n\> = \int d^4x\; e^{-iq\cdot x}\, \<n|j^\nu(x)\,j^\mu(0)|n\> .
\ee
Therefore
\ba
i\int d^4x\; e^{iq\cdot x}\, \<n|{\rm T}\left\{j^\nu(0)\,j^\mu(x)\right\}|n\> 
&=& i \int d^4x\; e^{iq\cdot x}\,\theta(x^0) \, \<n|[j^\mu(x),\, j^\nu(0)]|n\>
\\ && + i \int d^4x\; e^{-iq\cdot x}\, \<n|j^\nu(x)\,j^\mu(0)|n\>.
\nonumber
\ea
Recalling the definitions
\ba
G_T^{\mu\nu}(q) &=&  \frac{1}{Z} \sum_{n} e^{-\beta E_n} \,i\int d^4x\; e^{iq\cdot x}\, \<n|{\rm T}\left\{j^\nu(0)\,j^\mu(x)\right\}|n\> ,
\\
G_R^{\mu\nu}(q) &=&  \frac{1}{Z} \sum_{n} e^{-\beta E_n} \, i \int d^4x\; e^{iq\cdot x}\,\theta(x^0) \, \<n|[j^\mu(x),\, j^\nu(0)]|n\>,
\ea
we have the identity
\be
G_T^{\mu\nu}(q) = G_R^{\mu\nu}(q) + i \,\frac{\rho^{\nu\mu}(-q)}{1-e^{\beta q^0}}.
\ee
However, for $\mu=\nu$, the spectral function $\rho^{\nu\mu}$ is real, therefore 
\be\la{eq:GT_GR}
{\rm Re}(G_T^{\mu\nu}(q)) = {\rm Re}(G_R^{\mu\nu}(q))\qquad (\mu=\nu, \;q^0\in\mathbb{R}).
\ee

\subsection{The sum rules}

It follows from   Eq.\ (\ref{eq:GT_GR}) that 
\begin{equation}
\Re (G^{11}_R (q)) = \Re (G^{11}_T (q)) \quad q^0 \in \mathbb R \: ,
\end{equation}
and the equality also holds for $q^0=i\epsilon$.
In the rest frame of the fluid, and choosing the orientation $q = (q^0, 0, 0, \sqrt{(q^0)^2 + Q^2})$, we find that
\begin{equation}
\Re (H_R^{11} (q^0, Q^2;\;\hat z)) = \Re (\tilde F_1 (q^0, Q^2)) \: ,
\end{equation}
and Eq.~(\ref{eq:disp_integral}) reads in this case
\begin{equation}
\frac{1}{(m - 1)!} \biggl[ \frac{\dint^{m-1}}{\dint (\tilde q^0)^{m-1}} \: \Re ( \tilde F_1 (\tilde q^0, Q^2) ) \biggr]_{\tilde q^0 = i\epsilon} 
 =  \frac{1}{2 \pi} \int_{-\infty}^{\infty} \frac{\rho_1 (\tilde q^0, Q^2)}{(\tilde q^0-i\epsilon)^m} \: \dint \tilde q^0 \: ,
\end{equation}
where the functions $\rho_{1,2}$ are defined via a decomposition of the spectral function $\rho^{\mu \nu}$ analogous to Eq.~(\ref{eq:strf}),
see Eqs.~(\ref{eq:rho12}) and (\ref{eq:Wmunu_decomp}).
In the limit $Q \to \infty$, we can insert the expansion of Eq.~(\ref{eq:F1exp}).
Due to the fact that in the limit $\epsilon\to0$
\begin{equation}
\frac{1}{(m - 1)!} \biggl[ \frac{\dint^{m-1}}{\dint (q^0)^{m-1}} \: \bigl( \frac{2 T q^0}{Q^2}  \bigr)^n  \biggr]_{q^0 = i\epsilon} =
\biggl( \frac{2 T}{Q^2}  \biggr)^n \delta_{n, m-1} \: ,
\end{equation}
we find
\begin{equation}\la{eq:rho_sum_ruleq0}
\biggl( \frac{2 T}{Q^2}  \biggr)^n \sum_{f,j} 2 \: \mathcal Q_j^2 \: M_{fj} (Q, \tilde \mu) \langle O_{nj} \rangle = 
\frac{1}{\pi} \int_{0}^{\infty} \rho_1 (\tilde q^0, Q^2)\; \frac{(\tilde q^0)^{n+1}}{((\tilde q^0)^2+\epsilon^2)^{n+1}} \: \dint \tilde q^0 \: ,
\quad n = 2,4,\dots  \: .
\end{equation}
To rewrite the right-hand-side we have used the fact that $\rho_1 (q^0, Q^2)$ is odd in its first argument
and that $n+1$ is always an odd number in this context.
We now remark that if we first truncate $\rho_1$ to its leading-twist contribution,
the parameter $\epsilon$, which plays the role of an infrared regulator,
can be set to zero. This expectation is mathematically confirmed by the free-quark thermal spectral function;
and once only the leading-twist part is kept, partons with $x\gg 1$ are expected to be exponentially rare on physical grounds,
so that $(\rho_1)_{\textrm{leading-twist}}$ should fall off exponentially in $x$.
Proceeding in this way and 
changing the integration variable to $x = Q^2/(2 T q^0)$, we obtain the moment sum rule
\begin{equation}
\frac{1}{2 \pi} \int_0^{\infty} \dint x \: x^{n-1} [\rho_1 (x, Q^2)]_{\textrm{leading-twist}} = 
\sum_{f,j} \mathcal Q_f^2 M_{fj} (Q, \tilde \mu) \langle O_{nj} \rangle \: , \quad n = 2,4,\dots \: .
\label{eq:rho_sum_rule}
\end{equation}
Taking into account the relation (\ref{eq:rho12}), and considering
that $e^{- \beta q^0}$ is zero in the Bjorken limit,
Eq.~(\ref{eq:F1_sum_rule}) is recovered from the sum rule
Eq.~(\ref{eq:rho_sum_rule}). We remark that having to isolate the leading-twist part of the spectral function
in the dispersion integral distinguishes the thermal system from the case of the proton, for which
this operation is not necessary. This difference is related to the physical range of the variable $x$ being unbounded from above
in the thermal case.

\section{Verifying leading-twist predictions for the structure functions of the  plasma of non-interacting massless Dirac fermions} 
\label{sect:freeF1}

In this appendix, we explicitly verify the validity of the moment sum rules for non-interacting quarks using the leading-order Wilson coefficients,
i.e.\ Eq.\ (\ref{eq:F1_sum_rule}) with $M_{fj} \doteq \delta_{fj}$.

We note that the Euclidean-notation operator which coincides with $O_{nf}^{\mu_1\dots \mu_n}$ defined in Eq.\ (\ref{eq:Onf_def}) when 
all indices are temporal is 
\be
O_{{\rm E},nf}^{\mu_1\dots \mu_n} = \frac{1}{2} \left(\frac{-1}{2}\right)^{n-1}
{\cal S}\{\bar\psi_f \gamma^{\mu_1} \overleftrightarrow{D}^{\mu_2}\dots \overleftrightarrow{D}^{\mu_n} \psi_f \},
\ee
since $\frac{\partial}{\partial x^0} = i \frac{\partial}{\partial x_0^{\rm E}}$.

\subsection{The case $n=2$}
The twist-two, dimension-four operator reads
\be
O_{{\rm E},2f}^{\mu_1\mu_2} = \frac{-1}{4}\, \Big(\bar\psi_f \gamma^{\{\mu_1}\overleftrightarrow{D}^{\mu_2\}}\psi_f
 - \frac{1}{4}g^{\mu_1\mu_2} \bar\psi_f \overleftrightarrow{D}\!\!\!\!\!/ \;\psi_f\Big).
\ee
Now 
\be
\<O_{2f}^{00}\>_{\rm rest~frame} = -\frac{6T^4 N_c }{\pi^2} \sum_{n=1}^\infty \frac{(-1)^n}{n^4} = 
\frac{7\pi^2T^4 N_c }{120}.
\ee
Since $\<O_{2f}^{\mu_1\mu_2}\> = T^2 [u^{\mu_1}u^{\mu_2}- \frac{1}{4}g^{\mu_1\mu_2}]\<O_{2f}\>$,
we have for the RHS of Eq.\ (\ref{eq:F1_sum_rule})
\be\la{eq:eq:F1_sum_ruleRHS2}
\frac{1}{2}\<O_{2f}\> = \frac{2}{3T^2}\<O_{2f}^{00}\>_{\rm rest~frame} = \frac{7 \pi^2 T^2 N_c }{180}.
\ee 
Using Eq.\ (\ref{eq:freeF1}), 
one finds that the LHS of Eq.\ (\ref{eq:F1_sum_rule})  agrees with Eq.\ (\ref{eq:eq:F1_sum_ruleRHS2}).
We remark that the last member of Eq.\ (\ref{eq:eq:F1_sum_ruleRHS2}) is equal to $\frac{1}{4T^2}$ times 
the enthalpy density of $N_c$ massless Dirac fermions. This observation, given the interpretation (\ref{eq:F1f}) of the
structure function $F_1$ in terms of  a parton distribution function, confirms that the non-interacting quarks
carry the entire momentum $L^3(e+p)u$ of the fluid.

\subsection{The case $n=4$}
At tree-level, it suffices to write four permutations for the principal term (not containing the metric),
\ba
O_{E,4f}^{\mu_1\mu_2\mu_3\mu_4} &=& \frac{-1}{2} \frac{6}{4!\cdot8}\Big(\bar\psi_f \gamma^{\mu_1}\overleftrightarrow{D}^{\mu_2}
\overleftrightarrow{D}^{\mu_3}\overleftrightarrow{D}^{\mu_4}\psi_f + {\rm 3~perms}
\\ && - \frac{1}{4}\Big[g^{\mu_1\mu_2} \bar\psi_f \overleftrightarrow{D}\!\!\!\!\!/\;\overleftrightarrow{D}^{\mu_3}
\overleftrightarrow{D}^{\mu_4}\psi_f + {\rm perms} \Big]
\nonumber\\ && - \frac{1}{8} \Big[g^{\mu_2\mu_3}\bar\psi_f \gamma^{\mu_1}\overleftrightarrow{D}^2 D^{\mu_4}\psi_f + {\rm perms} \Big]
\nonumber\\ && + \frac{1}{12} \bar\psi_f\overleftrightarrow{D}\!\!\!\!\!/\; \overleftrightarrow{D}^2\psi_f 
\Big[g^{\mu_1\mu_2}g^{\mu_3\mu_4} + g^{\mu_1\mu_3} g^{\mu_2\mu_4} + g^{\mu_1\mu_4} g^{\mu_2\mu_3} \Big] \Big).
\nonumber
\ea
The coefficients of the trace terms have been determined in such a way that contracting any two indices of the expression
annihilates it.

The thermal expectation value of the zero-components is given by
\be
\<O_{4j}^{0000}\> = \frac{-1}{2} \< \bar\psi_f \gamma^0 (D^0)^3 \psi_f\>^T_{\rm vacuum}
= \frac{-120T^6 N_c}{\pi^2} \sum_{n=1}^\infty \frac{(-1)^n}{n^6} =  \frac{31\pi^4T^6N_c}{252}.
\ee
The reduced matrix element is determined by 
\ba
\<O_{4j}^{\mu_1\mu_2\mu_3\mu_4}\> &=& T^4 \Big(
 u^{\mu_1}u^{\mu_2}u^{\mu_3}u^{\mu_4}
 - \frac{1}{8} \Big[ g^{\mu_1\mu_2} u^{\mu_3} u^{\mu_4} + g^{\mu_1\mu_3} u^{\mu_2} u^{\mu_4} + \dots \Big]
\nonumber
\\ &&  + \frac{1}{48} \Big[ g^{\mu_1\mu_2}g^{\mu_3\mu_4}+ g^{\mu_1\mu_3}g^{\mu_2\mu_4} + g^{\mu_1\mu_4}g^{\mu_2\mu_3} \Big]\Big) \<O_{4j}\>.
\ea
Thus 
\be
\frac{1}{2}\<O_{4j}\> = \frac{16}{5T^4} \<O_{4j}^{0000}\> =   \frac{62\pi^4 T^2N_c}{315}.
\ee
One finds that this matches the LHS of Eq.\ (\ref{eq:F1_sum_rule}) for $n=4$.

\bibliography{dis_paper}

\end{document}